\documentclass[aps,11pt]{revtex4}
\usepackage{epsfig}
\usepackage{amsmath}
\usepackage{amsfonts}
\usepackage{bm}
\usepackage{times}
\usepackage{graphicx}

\newcommand{\be}{\begin{equation}}
\newcommand{\ee}{\end{equation}}
\newcommand{\ba}{\begin{eqnarray}}
\newcommand{\ea}{\end{eqnarray}}

\def\lp{\left(}
\def\rp{\right)}
\def\lc{\left[}
\def\rc{\right]}

\def\chip{\tilde{\chi}^+}
\def\chim{\tilde{\chi}^-}
\def\ma{m_{\chi_a}}
\def\mb{m_{\chi_b}}
\def\snu{\tilde{\nu}_e}
\def\ep{\epsilon}

\def\chin{\tilde{\chi}}

\def\DR{$\overline{DR}$}

\begin{document}
%
\title{NLO Polarized Chargino pair Production in Electron Positron 
Annihilation}
\author{Marco  A.  D\'{\i}az$^a$}
\author{Maximiliano A.   Rivera$^{b}$}
\author{Douglas A.  Ross$^c$}
\affiliation{
$^a$ Departamento de F\'\i sica, Pontificia
  Universidad  Cat\'{o}lica de Chile,  Santiago 690441,  Chile.\\ $^b$
  Departamento  de  F\'\i sica  y  Centro  de Estudios  Subat\'omicos,
  Universidad  T\'ecnica  Federico  Santa  Mar\'\i a,  Casilla  110-V,
  Valpara\'\i  so,  Chile.\\ $^c$  School  of  Physics and  Astronomy,
  University of Southampton, Highfield, Southampton SO17 1BJ, U.K.}
%
\begin{abstract}
We  calculate the complete  one-loop quantum  corrections to  the helicity
eigenstate  chargino  pair   production  cross  sections  in  polarized
electron positron collisions, within the Minimal  Supersymmetric 
Standard Model. We  calculate the 
non-QED   corrections using the helicity amplitudes formalism, and 
Dimensional  Regularization   to   deal  with
ultraviolet divergences.  We calculate QED corrections
using the dipole  subtraction formalism to
extract soft and collinear divergences in Bremsstrahlung, canceling 
them with the
infrared  divergences from virtual QED corrections. We  show numerical
results  for  the  Focus  Point  scenario in  mSUGRA,  where  we  find
important  quantum corrections  for differential  cross  sections with
definite chargino helicities.
\end{abstract}
\date{\today}
\maketitle
%
\section{Introduction}
%
In supersymmetry the fermionic partners of charged Higgs and $W$ gauge
bosons,  the  higgsinos and  winos,  mix to  form  a  pair of  charged
fermions called charginos, $\tilde \chi_i^\pm$ \cite{Haber:1984rc}. In
many supersymmetric scenarios they are light enough to be produced at
the  LHC   and  ILC,  although   they  have  not  yet   been  observed
\cite{Abbiendi:2003sc,Abdallah:2003xe,Heister:2002mn,Acciarri:1999km,
  Aaltonen:2008pv,Abazov:2009zi}.   The  discovery  of  heavy  charged
fermions   \cite{Polesello:2004aq,Blumenschein:2005ms}   not   coupled
through   strong   interactions  is   not   proof  of   supersymmetry,
though.   Precise   measurements  of   their   masses  and   couplings
\cite{Kalinowski:1998yn} should  exhibit the supersymmetric prediction
that SM  couplings have their mirror coupling  with superpartners. For
example, $WWZ$ coupling should be equal  to the $Z$ coupling to a pair
of winos.  This, together with the experimental  precision expected at
the  ILC,   leads  to  the  necessity  to   have  precise  theoretical
calculations  for the observables  in order  to properly  compare them
with the experimental results.

Quantum         corrections         to         chargino         masses
\cite{Pierce:1993gj,Eberl:2001eu,Fritzsche:2002bi,Schofbeck:2007ib},
chargino  production  cross section  in  electron positron  collisions
\cite{Diaz:1997kv,Ellis:1998jk,Oller:2005xg,Kilian:2006cj,Robens:2006np,
  Blank:2000uc,Kiyoura:1998yt},  chargino production cross  section in
hadron    colliders    \cite{Hao:2006df},    and    chargino    decays
\cite{Fujimoto:2007bn,Zhang:2001rd},  are  well  documented.  Here  we
improve  the collective  knowledge by  presenting a  complete one-loop
calculation to the production  cross section of polarized charginos in
collisions     between    polarized     electrons     and    positrons
\cite{Diaz:2001vm,Diaz:2000hi,Diaz:2002rr,Baer:2002bb}.     
We stress the fact that this is the first time complete NLO corrections
to the production of polarized charginos are calculated. These quantum
corrections are an effect of the rich electroweak properties of the
MSSM, and may provide a window to the Parity violating structure of
this supersymmetric model. The   full
potential  of  this calculation  will  become  available when  quantum
corrections  to   polarized  chargino  decays   are  incorporated,  as
indicated  by  the  tree-level  analysis  of the  importance  of  spin
correlation between chargino production and decay \cite{gudrid}.

In order to  handle the full one-loop calculation  we use form-factors
to  organize the bubble  and triangular  diagrams, and  $Q$-charges to
merge form-factors with box  diagrams. We use the  helicity amplitude
formalism \cite{Diaz:2001vm,Diaz:2000hi,Diaz:2002rr}  to keep track of
the   electron   and   positron   polarizations   and   the   chargino
helicities. To  calculate the QED corrections and  cancel the Infrared
(IR) divergences we use the dipole subtraction formalism introduced in
\cite{Dittmaier:1999mb}.  In  this   method  the  IR  divergences  are
isolated   analytically   through   the  introduction   of   auxiliary
subtraction  functions.   In  this  way, an IR cutoff is not  needed,
avoiding unpleasant  numerical problems that  emerge when a  cutoff is
used to regularize  the infinities.  At the end,  the three body phase
space integral is performed using a Monte Carlo method.
%
\section{Chargino Production at Tree Level}
\begin{figure}[!h]
\begin{center}
\includegraphics[angle=0,width=10cm]{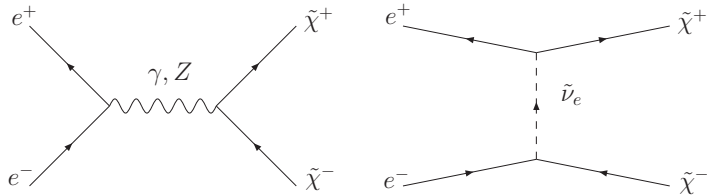}
\end{center}
\caption{Three   level   diagrams  for
  chargino pair productions at the ILC.\label{diagtreel}}
\end{figure}

%
In the  MSSM, the scattering $e^+e^-  \to \chip \chim$  arises at tree
level through $\gamma$ and $Z$ in the s-channel, and through $\snu$ in
the t-channel, as indicated in Fig.~\ref{diagtreel}.
%
%
In  this article we  include the  effects of  the polarization  of the
electron and  positron, and the chargino helicities.  The notation for
the four-momenta is as follows,
\begin{equation}
e^+(p_2)+e^-(p_1)\rightarrow \tilde{\chi}^+_b(k_2)+\tilde{\chi}^-_a(k_1)
\end{equation}
where the convention is that $p_1$ and $p_2$ are incoming and $k_1$ and 
$k_2$ are outgoing four-momenta. The $Z$-coupling to two electrons is, 
\begin{equation}
{\cal G}_{Zee}^{\mu, 0}
=-{g\over{2c_W}}\gamma^{\mu}\left[g^e_V-g^e_A\gamma_5\right]
\label{eq:ZeeTreeVertex}
\end{equation}
where $c_W=\cos\theta_W$, $s_W=\sin\theta_W$,  and $\theta_W$ the weak
mixing   angle,   with   $\sin^2\theta_W^{\overline{MS}}(m_Z)=0.23122$
\cite{Yao:2006px}.  Also $g^e_V=-1/2+2s_W^2$  and $g^e_A=-1/2$ are the
vector-axial couplings.   The $SU(2)$  gauge coupling constant  $g$ is
related  to   the  positron  electric   charge  by  $g=s_W   e$,  with
$\alpha(m_Z)=e^2/4\pi=1/127$  \cite{AguilarSaavedra:2005pw}.  We  also work
with  $m_Z=91.1876\pm 0.0021$  GeV and  $\Gamma_Z=2.4952\pm0.0023$ GeV
\cite{Abbiendi:2000hu}  for  the  mass  and  width of  the  $Z$  boson
respectively. The $Z$-coupling to a pair of charginos is,
\begin{equation}
{\cal G}_{Z\chi\chi}^{ab\,\nu, 0}={g\over{c_W}}\gamma^{\nu}\Big[
O'^L_{ab}\frac{1-\gamma_5}{2}+O'^R_{ab}\frac{1+\gamma_5}{2}\Big]
\label{eq:ZchichiTreeVertex}
\end{equation}
where we  use the notation in ref.~\cite{Haber:1984rc}  for the $O'_L$
and $O'_R$ couplings.

The photon couplings to a pair of electrons and a pair of charginos is
simply given by their electric charge,
\begin{equation}
{\cal G}_{\gamma ee}^{\mu, 0}=e\gamma^{\mu}\,,\qquad
{\cal G}_{\gamma\chi\chi}^{ab\,\nu, 0}=-e\gamma^{\nu}\delta_{ab}
\label{eq:GeeGchichiVertex}
\end{equation}
Note that our  convention is that a positive  chargino is the particle
and a negative chargino is the anti-particle.

The couplings between electrons charginos and sneutrinos are,
\begin{equation}
{\cal G}_{\tilde\nu_ee\chi}^{b+, 0}=-gV_{b1}\frac{1+\gamma_5}{2}C
\,,\qquad
{\cal G}_{\tilde\nu_ee\chi}^{a-, 0}=C^{-1}gV_{a1}\frac{1-\gamma_5}{2}
\label{eq:SneTreeVertex}
\end{equation}
where  the matrix  $V$ is  one of  the matrices  that  diagonalize the
chargino mass  matrix, in the notation of  \cite{Haber:1984rc}, and we
are assuming CP  is conserved and couplings are  real.  The matrix $C$
is  the charge  conjugation matrix,  and appears  due to  the clashing
arrows in the sneutrino  Feynman diagram in Fig.~\ref{diagtreel}. Only
experimental lower bounds for the sneutrino mass have been set, of 
which we mention
$m_{\tilde\nu_e}>94$ GeV \cite{Abdallah:2003xe}.
%
\section{Quantum Corrections}
%
We   regularize  divergent   diagrams   using  dimensional   reduction
$\overline{DR}$.   In each  graph,  divergences are  contained in  the
parameter
\begin{equation}
\Delta={2\over{4-n}}+\ln 4\pi-\gamma_E
\end{equation}
where  $\gamma_E$ is  the Euler-Mascheroni  constant, and  $n$  is the
number   of   space-time   dimensions,   which  is   defined   to   be
$n=4-2\epsilon$, with the limit $\epsilon\rightarrow 0$ to be taken at
the end  of the calculation.  We call  the renormalization subtraction
point $Q$, and it is such  that the parameters that define  the 
tree-level cross
section are promoted  to $\overline{DR}$ running parameters, evaluated
at the scale $Q$. We perform our calculations from first principles,
not relying on loop calculating packages, as a way to find an
independent result from groups that use these packages.

We have  two types of  divergences. The ultra-violet  (UV) divergences
appear at large internal momenta running in the loops, and they are to
be  cancelled  with  counterterms  introduced by  the  renormalization
procedure.   We also  have  diagrams with  infrared (IR)  divergences,
which  appear  at  low  internal  momenta. These  IR  divergences  are
cancelled  by bremsstrahlung,  i.e.,  real photon  emissions when  the
photon is soft or collinear.

We organize the calculation combining two formalisms, the form-factors
and the helicity $Q$-charges.  The form-factor formalism is specially
useful  to  organize  bubbles   and  triangle  graphs,  introduced  in
ref.~\cite{Diaz:1997kv}   in   the   context   of   the   quark-squark
approximation, in which only quark and squark loop corrections are 
taken into account.
The helicity  amplitude formalism  is  necessary when
boxes  are  included, and  was  introduced in  ref.~\cite{Diaz:2001vm}
where non-QED boxes were considered.  It is specially designed to keep
track  of the  polarization of  the  electrons and  positrons and  the
helicity of the charginos, in the context of full one-loop calculation
we  are dealing  with in  this paper.   We describe  and  combine both
formalisms below.
%
\subsection{Form-Factor Formalism}
%
In ref.~\cite{Diaz:1997kv} we  calculated the radiative corrections to
chargino production in electron-positron   collisions   in   the
approximation   where  only   loops  from   quark  and   squarks  were
considered.  A  form-factor  formalism  was developed,  and  here  we
generalize it to include triangular graphs in the gauge boson vertices
to electron-positron pairs as well.

We organize  the triangle  graphs with the  help of form-factors that
define Green's  functions where the fermionic lines  are on-shell, but
the bosonic line  are not. In the  case of $Z$ couplings to  a pair of
charginos, they are given by
\begin{equation}
{\cal G}_{Z\chi\chi}^{ab\,\mu}=
F_{Z\chi\chi}^{vR} \,\gamma^\mu \frac{(1+\gamma^5)}{2}
\, + \,   F_{Z\chi\chi}^{vL} \,\gamma^\mu \frac{(1-\gamma^5)}{2}
\, + \, F_{Z\chi\chi}^{sR} \,\frac{k_-^\mu}
{\sqrt{s}} \frac{(1+\gamma^5)}{2}
\, + \, F_{Z\chi\chi}^{sL} \,\frac{k_-^\mu}
{\sqrt{s}} \frac{(1-\gamma^5)}{2}
\end{equation}
and  similarly  for  ${\cal  G}_{\gamma\chi\chi}^{ab\,\mu}$  replacing
$Z\rightarrow\gamma$ in the form-factors. The $v$ and $s$ labels refer
to vector and scalar  couplings respectively. We define $k_-=k_1-k_2$,
which is  normalized by $\sqrt{s}$ in  order to make  the form-factors
dimensionless. Note that in ref.~\cite{Diaz:1997kv} we used a slightly
different but easily related definition for the form-factors.

In  the  case  of  $e^{\pm}$--sneutrino--chargino  vertices  the  
form-factors are,
\begin{equation}
{\cal G}_{\tilde\nu e\chi}^{b+}=
F_{\tilde{\nu}e\chi}^+ \frac{(1+\gamma^5)}{2} C \,,\qquad
{\cal G}_{\tilde\nu e\chi}^{a-}=
F_{\tilde{\nu}e\chi}^- C^{-1} \frac{(1-\gamma^5)}{2}
\label{Gnuechi}
\end{equation}
where $C$  is the charged conjugation  matrix. In eq.~(\ref{Gnuechi}),
the  $\pm$  notation  refers   to  the  positive  (negative)  outgoing
chargino.

Here  we extend  the formalism  in ref.~\cite{Diaz:1997kv}  to include
form-factors in the electron-positron vertex to a gauge boson,
\begin{equation}
{\cal G}_{Zee}^{\mu}=
F_{Zee}^{vR} \,\gamma^\mu \frac{(1+\gamma^5)}{2}
\, + \,   F_{Zee}^{vL} \,\gamma^\mu \frac{(1-\gamma^5)}{2}
\end{equation}
and similarly  for the  photon.  Only vector  form-factors  are needed
because scalar ones disappear  when electrons and positrons are taken
on-shell.  Also  note that in  the quark-squark approximation  used in
ref.~\cite{Diaz:1997kv} none of the $Zee$ form-factors receive quantum
corrections.

We list below the form-factors that do not vanish at tree-level,
\begin{eqnarray}
F_{Z\chi\chi}^{vR, 0}&{\hskip-1.65cm}=\frac{g}{c_W}O'^R_{ab}\,,\qquad 
&F_{Z\chi\chi}^{vL, 0}=\frac{g}{c_W}O'^L_{ab}
\nonumber\\
F_{\gamma\chi\chi}^{vR, 0}&{\hskip-1.8cm}=-e \delta_{ab}\,,\qquad 
&F_{\gamma\chi\chi}^{vL, 0}=-e \delta_{ab}
\nonumber\\
F_{\tilde\nu e\chi}^{+, 0}&{\hskip-1.75cm}=-g V_{b1}\,,\qquad &
F_{\tilde\nu e\chi}^{-, 0}=g V_{a1}
\\
F_{Zee}^{vR, 0}&=-\frac{g}{2c_W}(g^e_V-g^e_A)\,,\qquad 
&F_{Zee}^{vL, 0}=-\frac{g}{2c_W}(g^e_V+g^e_A)
\nonumber\\
F_{\gamma ee}^{vR, 0}&{\hskip-2.6cm}=e \,,\qquad 
&F_{\gamma ee}^{vL, 0}=e
\nonumber
\end{eqnarray}
We highlight a few details about these tree-level form-factors. First,
the  difference in  sign between  $F_{\gamma\chi\chi}$  and $F_{\gamma
  ee}$ is due to the fact we consider electrons and positive charginos
as particles, and positrons  and negative charginos as anti-particles,
and  of course $F_{\gamma\chi\chi}$  is proportional  to $\delta_{ab}$
due to electromagnetic  gauge invariance. Also, we note  that the 
form-factors $F_{\tilde\nu e\chi}$ indicate  that the wino component of the
chargino  is  the  one  that  couples to  the  sneutrino,  diminishing
considerably the $t$-channel contribution in the case of higgsino like
charginos. In contrast, since,
\begin{equation}
O'^R_{ab}=-U_{a1}U_{b1}-\textstyle{\frac{1}{2}}U_{a2}U_{b2}
+s_W^2\delta_{ab}
\,,\qquad
O'^L_{ab}=-V_{a1}V_{b1}-\textstyle{\frac{1}{2}}V_{a2}V_{b2}
+s_W^2\delta_{ab}
\,,
\end{equation}
we see  that both  the wino and  higgsino components of  the charginos
couple  to  the  $Z$  gauge   boson.   Finally,  we  note  that  since
$\sin^2\theta_W$  is  approximately equal  to  $1/4$,  the $Z$  vector
coupling $g^e_V$ to a pair of electrons is much smaller that the axial
one $g^e_A$. As a consequence, $F_{Zee}^{vR, 0}$ $F_{Zee}^{vL, 0}$ are
nearly  equal in  magnitude  but  with opposite  sign,  which in  turn
implies   a  cancellation   in   the  right   handed  electron   cross
sections. This  cancellation will become  more clear in  the following
section.

\subsection{Helicity Amplitudes Formalism}

The helicity amplitude formalism is a very systematic and economic way
to  organize virtual  corrections through  the so  called $Q$-charges,
introduced   in  \cite{Diaz:2001vm}.  Here   we  calculate   also  QED
corrections not included in the above reference.

Consider the scattering amplitude for electron-positron into a pair of
charginos.   On  the  one   hand,  the  electron  has  a  polarization
$\alpha=R,L$  and  momentum $p_1$,  while  the  positron has  opposite
polarization and  momentum $p_2$.  On  the other hand,  the negatively
charged  anti-chargino  has a  mass  $\ma$,  momentum  $k_1$, and  helicity
$\lambda_1$,  while the  positively charged  chargino has  a mass
$\mb$, a  momentum $k_2$, and helicity  $\lambda_2$. The dimensionless
scattering amplitudes are written as,
\begin{equation}
{\cal A}^\alpha_{\lambda_2,\lambda_1} \ = \ 
   \frac{1}{s} L_{\mu\alpha}\, 
    Q_{\alpha i}^\mu \, \langle k_2, \lambda_2 | \Gamma^i
 | k_1, \lambda_1 \rangle . \label{me1}
\end{equation}
where there is no sum on  $\alpha$. The normalization is such that the
differential scattering cross section is
\begin{equation}
\frac{d\sigma(\alpha,\lambda_2,\lambda_1)}{d\cos\theta}
  \ = \ \frac{\lambda^{1/2}(1,\ma^2/s,\mb^2/s)}{32 \, \pi \, s} 
 \left| {\cal A}^\alpha_{\lambda_2,\lambda_1} \right|^2,
\label{DiffCrossSect}
\end{equation}
and  $\lambda(x,y,z)\equiv   x^2+y^2+z^2-2xy-2xz-2yz$.   The  leptonic
matrix elements are,
\begin{equation}
L_{\mu\alpha}=\bar{v}(p_2)\gamma_\mu\frac{(1\pm \gamma^5)}{2}u(p_1)
\end{equation}
with the $+(-)$  sign for $\alpha=R(L)$.  Note that  there are no more
structures because  we are considering massless  leptons. The chargino
matrix elements are calculated with  the help of five Dirac structures
$\Gamma^i, i=1\cdots5$,
\begin{eqnarray} 
\Gamma^{1,2} & = & \frac{(1\pm\gamma^5)}{2} \nonumber  \\
\Gamma^{3,4} & = & \gamma^\nu\frac{(1\pm\gamma^5)}{2}  \\\nonumber 
\Gamma^5    & = & -i \, \sigma^{\nu\rho} 
\end{eqnarray}
Note that these structures  can have undisplayed Lorentz indices which
are  in  turn  contracted  with  correspondingly  undisplayed  Lorentz
indices  in  the   dimensionless  tensor  coefficients  $Q^\mu_{\alpha
  i}$. The  contribution from any  Feynman graph to such  an amplitude
can  always be  expressed  in this  form  by making  a suitable  Fierz
transformation where necessary. The tensor coefficients $Q^\mu_{\alpha
  i}$   can  be   further   reduced  to   scalar  $Q$-charges   ${\cal
  Q}^{\alpha}_i$,
\begin{eqnarray}
Q^\mu_{\alpha 1} & = & {\mathcal Q}^{\alpha}_1 \, k_-^\mu /\sqrt{s} \ 
\nonumber \\
Q^\mu_{\alpha 2} & = & {\mathcal Q}^{\alpha}_2 \, k_-^\mu /\sqrt{s} \ 
\nonumber \\
Q^{\mu\nu}_{\alpha 3} & = &{\mathcal Q}^\alpha_{3,1}\, g^{\mu\nu} \ \, + 
             {\mathcal Q}^\alpha_{3,2} \, k_-^\mu \,  p_-^\nu / s
\\
Q^{\mu\nu}_{\alpha 4} & = &{\mathcal Q}^\alpha_{4,1}\, g^{\mu\nu} \ \, + 
             {\mathcal Q}^\alpha_{4,2} \, k_-^\mu \, p_-^\nu / s
\nonumber \\
Q^{\mu\nu\rho}_{\alpha 5} 
& = &{\mathcal Q}^\alpha_{5,1}\, g^{\mu\nu} \,p_-^\rho/\sqrt{s} \ \, - 
\  i \,  {\mathcal Q}^\alpha_{5,2} \, \epsilon^{\mu\nu\rho\tau} \, 
p_{-\tau}/\sqrt{s},
\nonumber
\end{eqnarray}
where again, the scalar  $Q$-charges ${\cal Q}^{\alpha}_i$ may have an
extra  numeric  label,  and  they  are dimensionless.   In  the  above
equations    we   have    defined    $k_-^\mu=(k_1^\mu-k_2^\mu)$   and
$p_-^\mu=(p_1^\mu-p_2^\mu)$. Any  other structure can  be expressed in
terms  of  the above  quantities,  by  exploiting  the fact  that  the
leptonic  current  is  conserved  and  that  the  matrix  elements  of
$\Gamma^i$ are taken between on-shell chargino states.

In this article we concentrate on the case of chargino pair production
of equal  mass. We define  $k=|\vec k_1|=|\vec k_2|$ the  magnitude of
the  3-momentum  of  each  chargino  in  the  CM-frame,  $\theta$  the
scattering angle in the  same frame, and $v=2k/\sqrt{s}$ its velocity.
Helicity   amplitudes   are   given    for   the   general   case   in
ref.~\cite{Diaz:2001vm},  and   here  we   list  them  for   the  case
$m_{\chi_a}=m_{\chi_b}$.   Helicity amplitudes  are denoted  by ${\cal
  A}^{\alpha}_{\lambda_2\lambda_1}$,   where   $\alpha=L,R$   is   the
polarization  of  the   electron,  and  $\lambda_2\lambda_1$  are  the
helicities of  the chargino and anti-chargino  respectively.  For left
handed electrons we have:
\begin{eqnarray}
{\cal A}^L_{++}&=&
-{\cal Q}^L_1\,v\,(1-v) \sin\theta
+{\cal Q}^L_2\,v\,(1+v) \sin\theta 
\nonumber\\ &&
+({\cal Q}^L_{31}+{\cal Q}^L_{41})\sqrt{1 - v^2}\, \sin\theta
-({\cal Q}^L_{32}+{\cal Q}^L_{42})\,v\,\sqrt{1 - v^2}\,
\sin\theta\cos\theta 
\nonumber\\ &&
-2{\cal Q}^L_{51}\,v\,\sin\theta
+4{\cal Q}^L_{52}\,\sin\theta
\label{aLpp}
\\ & & \nonumber \\
{\cal A}^L_{+-}&=&
-{\cal Q}^L_{31}(1+v)\,(1+\cos\theta)
-{\cal Q}^L_{32}\,v\,(1+v)\,\sin^2\theta
\nonumber\\ &&
-{\cal Q}^L_{41}(1-v)\,(1+\cos\theta)
-{\cal Q}^L_{42}\,v\,(1-v)\,\sin^2\theta
\nonumber\\ &&
-4{\cal Q}^L_{52}\,\sqrt{1 - v^2}\,(1+\cos\theta)
\label{aLpm}
\\ & & \nonumber \\
{\cal A}^L_{-+}&=&
+{\cal Q}^L_{31}(1-v)\,(1-\cos\theta)
-{\cal Q}^L_{32}\,v\,(1-v)\,\sin^2\theta
\nonumber\\ &&
+{\cal Q}^L_{41}(1+v)\,(1-\cos\theta)
-{\cal Q}^L_{42}\,v\,(1+v)\,\sin^2\theta
\nonumber\\ &&
+4{\cal Q}^L_{52}\,\sqrt{1 - v^2}\,(1-\cos\theta)
\label{aLmp}
\\ & & \nonumber \\
{\cal A}^L_{--}&=&
-{\cal Q}^L_1\,v\,(1+v) \sin\theta
+{\cal Q}^L_2\,v\,(1-v) \sin\theta 
\nonumber\\ &&
-({\cal Q}^L_{31}+{\cal Q}^L_{41})\sqrt{1 - v^2}\, \sin\theta
+({\cal Q}^L_{32}+{\cal Q}^L_{42})\,v\,\sqrt{1 - v^2}\,
\sin\theta\cos\theta 
\nonumber\\ &&
-2{\cal Q}^L_{51}\,v\,\sin\theta
-4{\cal Q}^L_{52}\,\sin\theta
\label{aLmm}
\end{eqnarray}
where $v$ is the chargino velocity given by
\begin{equation}
v=\sqrt{1-{{4m_\chi^2}\over s}}
\end{equation}
The helicity amplitudes for right handed electrons are:
\begin{eqnarray}
{\cal A}^R_{++}&=&
-{\cal Q}^R_1\,v\,(1-v) \sin\theta
+{\cal Q}^R_2\,v\,(1+v) \sin\theta 
\nonumber\\ &&
+({\cal Q}^R_{31}+{\cal Q}^R_{41})\sqrt{1 - v^2}\, \sin\theta
-({\cal Q}^R_{32}+{\cal Q}^R_{42})\,v\,\sqrt{1 - v^2}\,
\sin\theta\cos\theta 
\nonumber\\ &&
+2{\cal Q}^R_{51}\,v\,\sin\theta
-4{\cal Q}^R_{52}\,\sin\theta
\label{aRpp}
\\ & & \nonumber \\
{\cal A}^R_{+-}&=&
+{\cal Q}^R_{31}(1+v)\,(1-\cos\theta)
-{\cal Q}^R_{32}\,v\,(1+v)\,\sin^2\theta
\nonumber\\ &&
+{\cal Q}^R_{41}(1-v)\,(1-\cos\theta)
-{\cal Q}^R_{42}\,v\,(1-v)\,\sin^2\theta
\nonumber\\ &&
-4{\cal Q}^R_{52}\,\sqrt{1 - v^2}\,(1-\cos\theta)
\label{aRpm}
\\ & & \nonumber \\
{\cal A}^R_{-+}&=&
-{\cal Q}^R_{31}(1-v)\,(1+\cos\theta)
-{\cal Q}^R_{32}\,v\,(1-v)\,\sin^2\theta
\nonumber\\ &&
-{\cal Q}^R_{41}(1+v)\,(1+\cos\theta)
-{\cal Q}^R_{42}\,v\,(1+v)\,\sin^2\theta
\nonumber\\ &&
+4{\cal Q}^R_{52}\,\sqrt{1 - v^2}\,(1+\cos\theta)
\label{aRmp}
\\ & & \nonumber \\
{\cal A}^R_{--}&=&
-{\cal Q}^R_1 \,v\,(1+v) \sin\theta
+{\cal Q}^R_2 \,v\,(1-v) \sin\theta 
\nonumber\\ &&
-({\cal Q}^R_{31}+{\cal Q}^R_{41})\sqrt{1 - v^2}\, \sin\theta
+({\cal Q}^R_{32}+{\cal Q}^R_{42})\,v\,\sqrt{1 - v^2}\,
\sin\theta\cos\theta 
\nonumber\\ &&
+2{\cal Q}^R_{51}\,v\,\sin\theta
+4{\cal Q}^R_{52}\,\sin\theta
\label{aRmm}
\end{eqnarray}
The  simplicity of  these expressions  is striking.   All  the quantum
corrections   are  concentrated   into  the   $Q$-charges.    To  find
differential cross sections for  specific chargino helicities, we just
have to  square the  corresponding amplitude above,  and feed  it into
eq.~(\ref{DiffCrossSect}).   If  we sum  the  squared amplitudes  over
chargino helicities we obtain from the expressions above the polarized
differential cross  sections when the beam is  $100\%$ polarized.  But
within  this approximation, we  do not  include spin  correlation. The
full potential of the result of this article will be realized once the
decay  chain  is  added,  calculating the  chargino  spin  correlation
between  production  and  decay   \cite{gudrid},  and  its  effect  on
observables like angular distributions of final decay products.

It  is instructive to  analyze the  tree-level approximation  with the
help  of   the  $Q$-charges.   The  tree-level  expressions   for  the
$Q$-charges are,
\begin{eqnarray}
{\cal Q}^{L,0}_{31}&=&
-\frac{g^2}{2c_W^2}(g^e_V+g^e_A)\frac{s}{s-m_Z^2}O'^R_{ab}
-e^2\delta_{ab}
\nonumber\\
{\cal Q}^{L,0}_{41}&=&
-\frac{g^2}{2c_W^2}(g^e_V+g^e_A)\frac{s}{s-m_Z^2}O'^L_{ab}
-e^2 \delta_{ab}
-g^2 V_{b1}\frac{s}{t-m^2_{\tilde\nu}}V_{a1}
\nonumber\\
{\cal Q}^{R,0}_{31}&=&
-\frac{g^2}{2c_W^2}(g^e_V-g^e_A)\frac{s}{s-m_Z^2}O'^R_{ab}
-e^2 \delta_{ab}
\label{QtreeLevel}\\
{\cal Q}^{R,0}_{41}&=&
-\frac{g^2}{2c_W^2}(g^e_V-g^e_A)\frac{s}{s-m_Z^2}O'^L_{ab}
-e^2 \delta_{ab}
\nonumber
\end{eqnarray}
This     coincides      with     the     expressions      given     in
ref.~\cite{Kalinowski:1998yn}  (after allowing  for the  fact  that in
ref.~\cite{Kalinowski:1998yn} the negatively charged chargino is taken
to be  the particle and  the positively charged one  the antiparticle,
whereas  our convention  is {\it  vice  versa}).  Note  that the  only
dependence on the scattering angle $\theta$ of the $Q$-charges at tree
level  comes  from  the  $t$-Mandelstam variable  from  the  sneutrino
contribution, which is always negative.

With the help  of eq.~(\ref{QtreeLevel}) it is easy  to understand why
the right  handed electron  cross sections are  smaller than  the left
handed ones. Since $g^e_V$ is  very small and $g^e_A$ is negative, the
coefficients  for  the $O'$  couplings  are  positive  for left  handed
electrons and negative  for right handed ones. On  the other hand, for
equal charginos the $O'$ couplings reduce to,
\begin{equation}
O'^R_{11}=\textstyle{\frac{1}{2}}\sin^2\phi_R-c^2_W
\,,\qquad
O'^L_{11}=\textstyle{\frac{1}{2}}\sin^2\phi_L-c_W^2
\end{equation}
where $\phi_R$ is the rotation  angle that defines the rotation matrix
$U$, and  analogously for  the angle $\phi_L$  and the matrix  $V$. In
this way, the $O'_{aa}$ are  always negative, implying that the photon
and  $Z$  contributions   interfere  destructively  for  right  handed
electrons and constructively for left handed ones.

\section{Non-QED Virtual Corrections}
\label{subsub:mssm}

In this article we  present separately QED corrections (bremsstrahlung
and virtual photonic graphs), from the rest of the virtual graphs.
The later are analyzed in  this section. We distinguish three kinds of
diagrams  which contribute  to the  total cross  section:  bubble (two
point functions),  triangular (three  point functions), and  box (four
point functions) diagrams. We work in the Feynman gauge $\xi=1$.

\begin{figure}[!h]
\begin{center}
\includegraphics[angle=0,width=5cm]{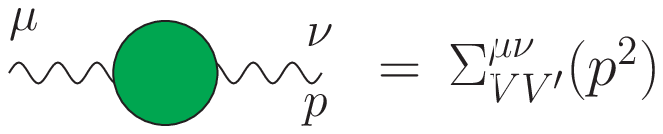}
\end{center}
\caption{Vector boson two-point Green function.
\label{self:v}}
\end{figure}

Bubble diagrams can  be easily inserted into the  form-factors defined
before.  We  consider first the  two-point Green functions  with gauge
bosons in the  external legs. The two gauge bosons,  which we call $V$
and $V'$,  are off-shell,  and in our  case they correspond  either to
neutral $Z$ or  $\gamma$. In Fig.~\ref{self:v} we see  the diagram for
this Green function,
%
whose  Lorentz  structure is  defined  with  the  help of  two  scalar
functions $A$ and $B$,
\begin{equation}
\Sigma_{VV'}^{\mu\nu}(p^2) =
i[A_{VV'}(p^2)g^{\mu \nu}+B_{VV'}(p^2)p^{\mu}p^{\nu}],
\label{prop:vector}
\end{equation}
Complete prototype for bubble  diagrams will be given elsewhere, while
the $A$ function can be found in \cite{Diaz:2001vm}.

The photon  self-energy vanishes at  zero momentum by virtue  of gauge
invariance  so that  after subtraction  in the  $\overline{DR}$ scheme
contributes to the photon form-factor according to
\begin{equation}
\Delta F_{\gamma\chi\chi}^{vR(L)}=
-e{{A_{\gamma\gamma}(s)}\over{s}}\delta_{ab}
\end{equation}
where $a$ and $b$ refer to the two species of charginos produced.  The
photon-$Z$ mixing is also subtracted in the $\overline{DR}$ scheme and
contributes to the $Z$ form-factor
\begin{equation}
\Delta F_{Z\chi\chi}^{vR(L)}=
-e{{A_{Z\gamma}(s)}\over{s}}\delta_{ab}
\end{equation}
and to photon form-factors:
\begin{equation}
\Delta F_{\gamma\chi\chi}^{vR(L)}=
{g\over{c_W}}O'^{R(L)}_{ab}{{A_{\gamma Z}(s)}\over{s-m_Z^2}}
\end{equation}
The  $Z$--boson  self-energy  is  regularized with  a  subtraction  at
$s=m_Z^2$:
\begin{equation}
\Delta F_{Z\chi\chi}^{vR(L)}=
{g\over{c_W}}O'^{R(L)}_{ab}{{A_{ZZ}(s)-
A_{ZZ}(m_Z^2)}\over{s-m_Z^2}}
\end{equation}

\begin{figure}[!h]
\begin{center}
\includegraphics[angle=0,width=5cm]{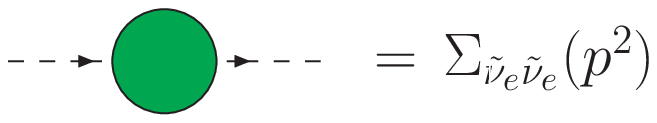}
\end{center}
\caption{Sneutrino self energy.\label{self:s}}
\end{figure}

Another  important  two-point Green  function  is  the sneutrino  self
energy, whose diagram is in Fig.~\ref{self:s}
Since it  is Lorentz  scalar, it is  represented by only  one function
$A$,
\begin{equation} 
\Sigma_{\snu \snu}(p^2)=i[A_{\snu \snu}(p^2)].
\end{equation}
The  contribution to  the  sneutrino form-factors  is obtained  after
regularizing  with  a subtraction  at  $t=m_{\tilde\nu}^2$, with  the
following result,
\begin{equation}
\Delta F_{\tilde\nu e\chi}^{\pm}=
{1\over2}gV_{b(a)1}{{A_{\tilde\nu\tilde\nu}(t)-
A_{\tilde\nu\tilde\nu}(m_{\tilde\nu}^2)}\over{t-m_{\tilde\nu}^2}}
\end{equation}
This  guarantees   that  the  parameters   $m_Z$  and  $m_{\tilde\nu}$
respectively refer to the physical (pole-)masses.

\begin{figure}[!h]
\begin{center}
\includegraphics[angle=0,width=5cm]{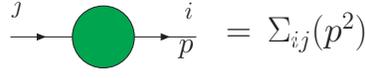}
\end{center}
\caption{Chargino two-point Green function.\label{self:f}}
\end{figure}

The chargino two-point function can be seen in Fig.~\ref{self:f}
where the  external legs correspond to  any of the  two charginos. Its
Lorentz decomposition involves four scalar functions,
\begin{equation} 
\Sigma_{ij}(p)  =
i\left[A^+_{ij}(p^2)+B^+_{ij}(p^2)\,\,\slash
{\!\!\! p}\right]\frac{(1+\gamma_5)}{2}
+
i\left[A^-_{ij}(p^2)+B^-_{ij}(p^2)\,\,\slash
{\!\!\! p}\right]\frac{(1-\gamma_5)}{2}.
\end{equation}
The chargino self-energy and  mixing contribute to form-factors  in a more
complicated way.  Since these are external particles  we have insisted
that the subtractions are performed on-shell, so that the renormalized
chargino fields  are indeed physical  fields. Details can be  found in
ref.~\cite{Diaz:2001vm,Diaz:1997kv}.

Also embedded into form-factors are triangular diagrams, grouped into
$Z\chi\chi$, $Zee$, $\gamma\chi\chi$, $\gamma ee$ and $e\chi\tilde\nu$
three-point functions.  Form-factors  are in turn  easily incorporated
into $Q$-charges.  On the one  hand, the left handed  $Q$-charges that
receive contributions from form-factors are,
\begin{eqnarray}
\Delta Q^L_1&=&-\frac{g}{2c_W}(g^e_V+g^e_A)
\frac{s}{s-m_Z^2}\Delta F^{sR}_{Z\chi\chi}
+e\Delta F^{sR}_{\gamma\chi\chi}
\nonumber\\
\Delta Q^L_2&=&-\frac{g}{2c_W}(g^e_V+g^e_A)
\frac{s}{s-m_Z^2}\Delta F^{sL}_{Z\chi\chi}
+e\Delta F^{sL}_{\gamma\chi\chi}
\\
\Delta Q^L_{31}&=&-\frac{g}{2c_W}(g^e_V+g^e_A)
\frac{s}{s-m_Z^2}\Delta F^{vR}_{Z\chi\chi}
+\frac{g}{c_W}O'^R_{ab}\frac{s}{s-m_Z^2}\Delta F^{vL}_{Zee}
+e\Delta F^{vR}_{\gamma\chi\chi}
-e\delta_{ab}\Delta F^{vL}_{\gamma ee}
\nonumber\\
\Delta Q^L_{41}&=&-\frac{g}{2c_W}(g^e_V+g^e_A)
\frac{s}{s-m_Z^2}\Delta F^{vL}_{Z\chi\chi}
+\frac{g}{c_W}O'^L_{ab}\frac{s}{s-m_Z^2}\Delta F^{vL}_{Zee}
+e\Delta F^{vL}_{\gamma\chi\chi}
-e\delta_{ab}\Delta F^{vL}_{\gamma ee}
\nonumber\\
&&-\frac{1}{2}gV_{b1}\frac{s}{t-m^2_{\tilde\nu}}\Delta F^-_{\tilde\nu e\chi}
+\frac{1}{2}gV_{a1}\frac{s}{t-m^2_{\tilde\nu}}\Delta F^+_{\tilde\nu e\chi}
\nonumber
\end{eqnarray}
On   the  other   hand,   the  right   handed  $Q$-charges   receiving
contributions from form-factors are given by,
\begin{eqnarray}
\Delta Q^R_1&=&-\frac{g}{2c_W}(g^e_V-g^e_A)
\frac{s}{s-m_Z^2}\Delta F^{sR}_{Z\chi\chi}
+e\Delta F^{sR}_{\gamma\chi\chi}
\nonumber\\
\Delta Q^R_2&=&-\frac{g}{2c_W}(g^e_V-g^e_A)
\frac{s}{s-m_Z^2}\Delta F^{sL}_{Z\chi\chi}
+e\Delta F^{sL}_{\gamma\chi\chi}
\\
\Delta Q^R_{31}&=&-\frac{g}{2c_W}(g^e_V-g^e_A)
\frac{s}{s-m_Z^2}\Delta F^{vR}_{Z\chi\chi}
+\frac{g}{c_W}O'^R_{ab}\frac{s}{s-m_Z^2}\Delta F^{vR}_{Zee}
+e\Delta F^{vR}_{\gamma\chi\chi}
-e\delta_{ab}\Delta F^{vR}_{\gamma ee}
\nonumber\\
\Delta Q^R_{41}&=&-\frac{g}{2c_W}(g^e_V-g^e_A)
\frac{s}{s-m_Z^2}\Delta F^{vL}_{Z\chi\chi}
+\frac{g}{c_W}O'^L_{ab}\frac{s}{s-m_Z^2}\Delta F^{vR}_{Zee}
+e\Delta F^{vL}_{\gamma\chi\chi}
-e\delta_{ab}\Delta F^{vR}_{\gamma ee}
\nonumber
\end{eqnarray}
Finally, box diagrams are  directly incorporated into $Q$-charges with
prototype  diagrams  in   ref.~\cite{Diaz:2001vm},  and  more  general
diagrams that will be shown elsewhere.

As opposed to  bubbles and some triangular diagrams,  box diagrams are
UV finite. Ultraviolet divergences that occur in a few of the triangle
graphs are  subtracted in the $\overline{DR}$  scheme with subtraction
point  $Q$. Therefore,  apart for  the masses  which are  taken  to be
physical and the weak-mixing  angle whose renormalization is described
above,  all  other  parameters are  to  be  considered  to be  in  the
$\overline{DR}$ at the scale $Q$  {\it in the MSSM theory}. This means
that  the  translation of  the  values  used  here to  those  directly
extracted from experiment, such as neutral current neutrino scattering
cross-sections  or  the   measured  fine-structure  constant  will  be
slightly  different  from that  of  the  Standard  Model (without  the
supersymmetric partners). For example, the treatment of the photon-$Z$
propagator  system, described above,  guarantees that  the propagators
only have poles at zero and  $M_Z$, but there is still some remnant of
photon-$Z$ mixing at these poles. We have checked numerically that the
effect of a further subtraction of the photon-$Z$ mixing propagator to
remove  this   mixing  has  a  negligible  numerical   effect  on  our
results.  Furthermore, the  input SUSY  parameters chosen  are assumed
also to be the corresponding values renormalized in this scheme at the
same  scale. We  expect  the sensitivity  to  (reasonable) changes  in
renormalization scheme to be  genuinely of order $\alpha_W/\pi$ and to
have no significant effect on our numerical results. The numerical value
we use for the subtraction scale in this article is $Q=1$ TeV as 
suggested by the SPA Project \cite{AguilarSaavedra:2005pw}. 

\section{QED Corrections}
\label{subsub:qed}

In the  previous section we discussed the  non-QED virtual corrections
to  our chargino  production process.   In this  section we discuss the
treatment of the remaining  QED corrections  separately  
because special  care must  be
taken  with IR divergences  present in  bubbles, triangles,  and boxes
involving photons. These IR  divergences appear in graphs with virtual
photons at low momenta of  internal particles in the loop. They cancel
from the cross section  with corresponding IR divergences comming from
real  photon   emission  from  the  external   fermions,  {\it  i.e.},
bremsstrahlung  \cite{Bloch:1937pw,Kinoshita:1962ur,Lee:1964is}, which
appear  in both  soft and  collinear  photons.  As  a remanent,  large
corrections proportional
to  $\ln(s/m^2_e)$  remain, known  as  Leading  Logarithms (these  are
initial state radiation corrections) and  common to any of such similar
process. For bremsstrahlung corrections  we do not use the $Q$-charges
formalism,  calculating  directly  the  corrections to  the  amplitude
squared,      using      the      algebraic      manipulator      FORM
\cite{Vermaseren:2000nd}.

\subsection{Bremsstrahlung corrections}

\begin{figure}[!h]
\begin{center}
\includegraphics[angle=0,width=13.5cm]{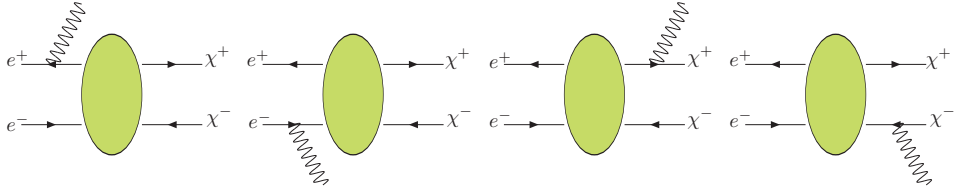}
\end{center}
\caption{Real photon emission.
\label{bremss}}
\end{figure}

Diagrams contributing to the process $e^+e^-\to\chip\chim\gamma$ are 
depicted in Fig.~\ref{bremss},
where  the bubbles  represent the  three different  tree-level channels
($\gamma$, $Z$, and $\tilde{\nu}_e$).

The total cross section $\sigma_{real}$ for a real photon emission can
be written as follow,
\begin{equation}
\sigma_{real}=\int d\Phi_3\,\sum_{\lambda_{\gamma}}\,|\mathcal{M}_{real}|^2,
\label{eq:real:cs}
\end{equation}
where $d\Phi_3$  is the three-body  phase space
and $\lambda_{\gamma}$  is the  photon polarization. We  implement the
dipole subtraction formalism  introduced in \cite{Dittmaier:1999mb} in
order  to analytically  isolate  the IR  divergences  coming from  the
emission of real photons, and  cancel them from the also analytically
extracted   divergences  from   virtual  photon   diagrams.   Once  IR
divergences  are removed,  the QED  corrections  are IR  free and  are
calculated    numerically.    The   dipole    subtraction   formalism
conveniently avoids  numerical problems arising when a  cutoff is used
to regularize the infinities.

The dipole subtraction method entails the introduction of an \emph{ad
  hoc} auxiliary function  $|\mathcal{M}_{sub}|^2$ which becomes equal
to  $\sum_{\lambda_{\gamma}}\,|\mathcal{M}_{real}|^2$ in the  soft and
collinear limits,
\begin{equation}
|{\cal M}_{sub}|^2 \rightarrow
 \sum_{\lambda_{\gamma}} 
|\mathcal{M}_{real}|^2, \ \ \mbox{as} \ \, \quad p_i\cdot k \rightarrow 0,
\end{equation}
where $p_i$ are the four-momenta of the massless external fermions and
$k$  is the  four-moment of  the photon.  Adding and  subtracting this
function into eq.~(\ref{eq:real:cs}) we find,
\begin{equation}
\sigma_{real}
= \int d\Phi_3\, ( \sum_{\lambda_{\gamma}}\,|\mathcal{M}_{real}|^2-
|\mathcal{M}_{sub}|^2 )  + \int d\Phi_3\, |\mathcal{M}_{sub}|^2.
\label{eq:real}
\end{equation}
where  the  first integral  is  finite  and  is performed  numerically
without the serious instabilities  found in cutoff methods. The second
integral  can  be  done  after  an analytical  extraction  of  the  IR
divergences, regularized  in \cite{Dittmaier:1999mb} by  introducing a
non-zero photon mass $m_\gamma$, and an electron mass $m_e$.  Once the
(universal)  contribution  proportional  to  $\ln(m_e)$  arising  from
initial-state radiation  is removed, the soft  and collinear divergent
terms  can be  mapped into  pole terms  in  dimensional regularization
using the mappings
\begin{eqnarray}
 \ln \frac{m^2_{\gamma}}{Q^2}=\ln \frac{m^2_{e}}{Q^2} & \to &
 \frac{1}{\epsilon}
\nonumber \\ 
 \ln \frac{m^2_{\gamma}}{Q^2}\ln \frac{m^2_{e}}{Q^2}-
\frac{1}{2}\ln^2 \frac{m^2_{e}}{Q^2} & \to & \frac{1}{\epsilon^2}
\label{mapping}
\end{eqnarray}
These  IR  divergences  cancel  against IR  divergences  from  virtual
diagrams (also treated by dimensional regularization).

With   the  process   described   above,  the   IR  divergences   from
bremsstrahlung diagrams are,
\begin{equation}
\sigma^{\alpha\lambda_1\lambda_2}_{real,\, \textnormal{\scriptsize{ IR}}}
=\frac{\alpha_e}{2\pi}
\lc \frac{2}{\ep^2} + \frac{5}{\ep}+\frac{2}{\ep} \ln \lp \frac{Q^2}{s}
\rp +\frac{2}{\ep}\frac{1+v^2}{v}\ln \frac{1-v}{1+v}
+\frac{4}{\ep}\ln\frac{m_{\chin}^2-u}{m_{\chin}^2-t}\rc
\sigma_{2\to 2}^{\alpha\lambda_1\lambda_2}
\label{real:dsigm:fin}
\end{equation}
which  as we  see, are  proportional to  the tree-level  chargino pair
production cross section.  In the above result, the  double pole comes
from simultaneous soft and  collinear divergences, the simple pole not
proportional to any logarithm  comes from collinear divergences, while
the   simple  pole   proportional  to   logarithms  comes   from  soft
divergences.
%
\subsection{Virtual Corrections}
%
Virtual QED quantum corrections at NLO in $\alpha_e$ are given by,
\begin{equation}
\sigma^{\alpha\lambda_1\lambda_2}_{virt}=\frac{\alpha_e}{4\pi}
\int d\Phi_2 \ 2\mathbb{R}e\langle(
{\cal M}_{2\to 2}^{\alpha\lambda_1\lambda_2})^*
{\cal M}_{virt}^{\alpha\lambda_1\lambda_2} \rangle 
\label{cross:qed}
\end{equation}
where ${\cal M}_{2\to 2}^{\alpha\lambda_1\lambda_2}$ is the tree level
amplitude,  ${\cal M}_{virt}^{\alpha\lambda_1\lambda_2}$  contains all
QED virtual corrections, and the photon coupling to fermions is factor
out in $\alpha_e/(4\pi)$. In  \DR, the IR divergent contributions from
virtual QED correction can be written as,
\begin{equation}
\sigma^{\alpha\lambda_1\lambda_2}_{virt,\, \textnormal{\scriptsize{ IR}}}
= -\frac{\alpha_e}{2\pi}
\lc \frac{2}{\ep^2} + \frac{5}{\ep}+\frac{2}{\ep} \ln \lp \frac{Q^2}{s}
\rp +\frac{2}{\ep}\frac{1+v^2}{v}\ln \frac{1-v}{1+v}
+\frac{4}{\ep}\ln\frac{m_{\chin}^2-u}{m_{\chin}^2-t}\rc
\sigma_{2\to 2}^{\alpha\lambda_1\lambda_2}
\label{virt:dsigma:fin}
\end{equation}
where    details    will   be    given    elsewhere.   Clearly    from
eqs.~(\ref{real:dsigm:fin})  and (\ref{virt:dsigma:fin}),  virtual and
real IR divergences cancel each other.
%
\section{Numerical Results}
%
 As a working  example we concentrate on the  mSUGRA Focus Point known
 as SPS2 \cite{Allanach:2002nj}.  This scenario is characterized by,
\begin{center}
\begin{tabular}{ccccc}
$m_0=1450$ GeV, & $m_{1/2}= 300$ GeV,  & $A_0= 0$ GeV, & $\tan \beta =
  10$, & $\mu>0$
\end{tabular}
\end{center}
and we use the code ISAJET for the running from GUT to weak scale
\cite{Baer:1999sp}. The
low energy soft parameters calculated  this way are fed into our code.
The  integration  over phase  space  is  performed  with a  MonteCarlo
technique.  Some  relevant low energy parameters and  masses are shown
in the following table,
\begin{center}
\begin{tabular}{|c|c|}
\hline
parameters (GeV) & masses (GeV) \\ \hline
$M_1= 121.9$      & $m_{\chin_1^+}^0= 172.1$  \\
$M_2= 235.7$      & $m_{\chin_2^+}^0= 297.0$  \\
$\mu=  222.2$  & $m_{\chin^0_1}=111.5$     \\
               & $m_{\snu}= 1454$         \\
\hline
\end{tabular}
\end{center}
where  the charginos, neutralino,  and sneutrino  masses are  given at
tree-level.  We correct the chargino  masses at one-loop, and in
order  to  compare the  corrected  cross  section  without moving  the
threshold, we define  a tree-level scenario where the  values of $M_2$
and $\mu$ are tuned such that  this tree-level masses are equal to the
one-loop corrected  chargino masses.  These  new tree-level parameters
and one-loop corrected chargino masses are in the following table,
\begin{center}
\begin{tabular}{|c|c|}
\hline
parameters (GeV)     & masses (GeV)          \\ \hline
$M_2^0= 242.9$       & $m_{\chin_1^+}= 173.7$  \\
$\mu^0= 219.7$       & $m_{\chin_2^+}= 300.0$  \\
\hline
\end{tabular}
\end{center}
We report below only on  production cross sections of light charginos,
with a  mass of $m_{\chi^+_1}=173.7$ GeV  as shown on  the table.

As we mentioned before, the renormalization scale we use is $Q=1$ TeV
motivated by the SPA Project \cite{AguilarSaavedra:2005pw}. Electroweak 
observables $\sin^2\theta_W^{\overline{MS}}(m_z)$ and $\alpha(m_Z)$ 
mentioned in section II must be run up to the scale $Q=1$ TeV. We
find $\sin^2\theta_W^{\overline{MS}}(Q)=0.2395$ and $\alpha(Q)=1/125$,
which translates into $g(Q)=0.6479$ and $g'(Q)=0.3636$, as reported
in \cite{AguilarSaavedra:2005pw}.

We  separate  the  corrections  in  the  following  way:  by  ``MSSM''
corrections we mean all loops  that do not include photons, by ``QED''
corrections we mean all loops including photons (virtual QED) plus all
bremsstrahlung  diagrams, and  by ``$\ell\ell$''  we mean  the leading
logarithms, which are universal to all this kind of processes.

Our definition for the ``$\ell\ell$'' contribution is,
\begin{equation}
\sigma_{\ell\ell}^{\alpha\lambda_1\lambda_2}=
\frac{\alpha}{\pi}\ln\left(\frac{s}{m_e^2}\right)
\int_0^1\,dx\,\frac{1+x^2}{1-x}
\left[\int d\Phi_2(x)
\Big|{\cal M}_{2\rightarrow 2}^{\alpha\lambda_1\lambda_2}(xp)\Big|^2
-\int d\Phi_2(1)
\Big|{\cal M}_{2\rightarrow 2}^{\alpha\lambda_1\lambda_2}(p)\Big|^2
\right] \label{lleq}
\end{equation}
where  the  term  $\ln\left(\frac{s}{m_e^2}\right)$ is  precisely  the
large logarithm  that gives  the name to  the whole  leading logarithm
contribution.  The function $(1+x^2)/(1-x)$ is the splitting function,
which  is related  to  the probability  of  finding an  electron or  a
positron with a  momentum $xp$ before the emission  of a photon, which
takes  a  momentum $(1-x)p$.   The  differential  $d\Phi_2(x)$ is  the
2-body  phase space with  one incoming  4-momentum multiplied  by $x$,
which translates into $s\rightarrow xs$.
%
\subsection{Unpolarized Total Cross Section}
%

%
\begin{figure}[!h]
\begin{center}
\includegraphics[angle=0,width=12cm]{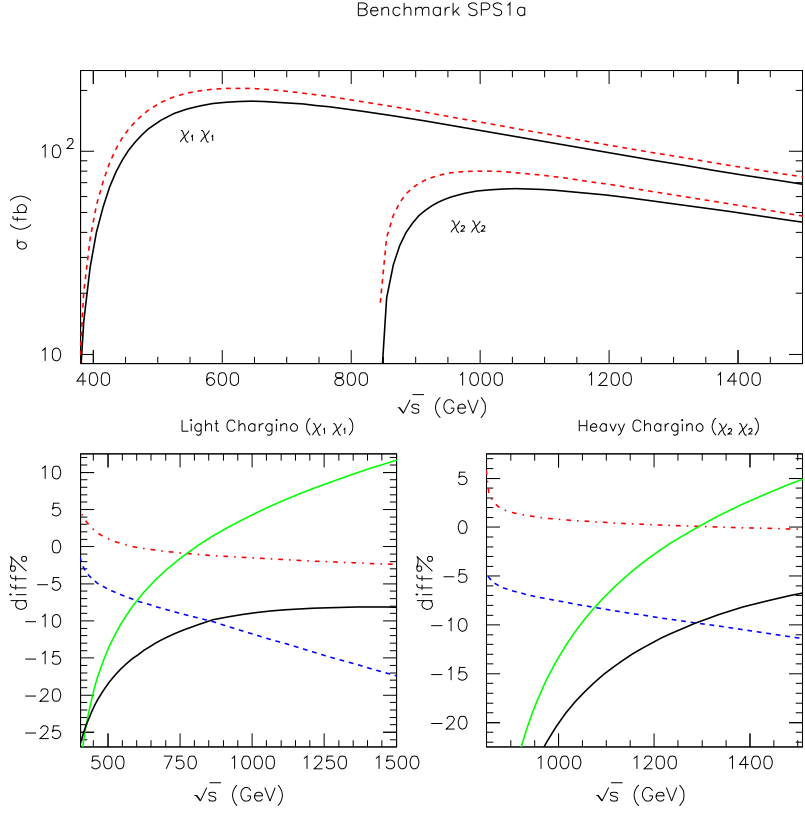}
\end{center}
\caption{Unpolarized   total   cross   section   as  a   function   of
  $\sqrt{s}$ for the SPS1a scenario. In the upper frame we compare the 
tree-level (red-dashed), with full NLO (black-solid). In the lower 
frames, the different curves correspond to full NLO corrected 
(black solid), MSSM (blue dashes), QED (red dot-dash), and  $\ell\ell$ 
(green solid).
\label{sps1a_runs}}
\end{figure}
%
We start by comparing our unpolarized results with the ones reported in 
\cite{Oller:2005xg}. In this reference the result for NLO corrections
to the unpolarized cross section was reported for the scenario known as 
SPS1a. In that case, complete NLO quantum corrections are very large and 
negative, strengthen by a large contribution from sneutrinos, due to a 
small sneutrino mass and large sneutrino-neutrino-electron coupling. 
Our predictions for the total NLO cross section for light charginos in 
SPS1a, whose corrections vary from $-27\%$ at $\sqrt{s}=400$ GeV, to 
$-7\%$ at $\sqrt{s}=1400$ GeV, are in reasonable agreement with the ones 
reported in \cite{Oller:2005xg} within $\pm 2\%$, as can be seen in the 
lower-left frame of our Fig.~\ref{sps1a_runs}. Similarly, the total NLO 
cross section for heavy charginos vary from $-22\%$ at $\sqrt{s}=980$
GeV to $-8\%$ at $\sqrt{s}=1400$ GeV, and they are also in agreement 
within $\pm 2\%$ with \cite{Oller:2005xg}.

\begin{figure}[!h]
\begin{center}
\includegraphics[angle=0,width=12cm]{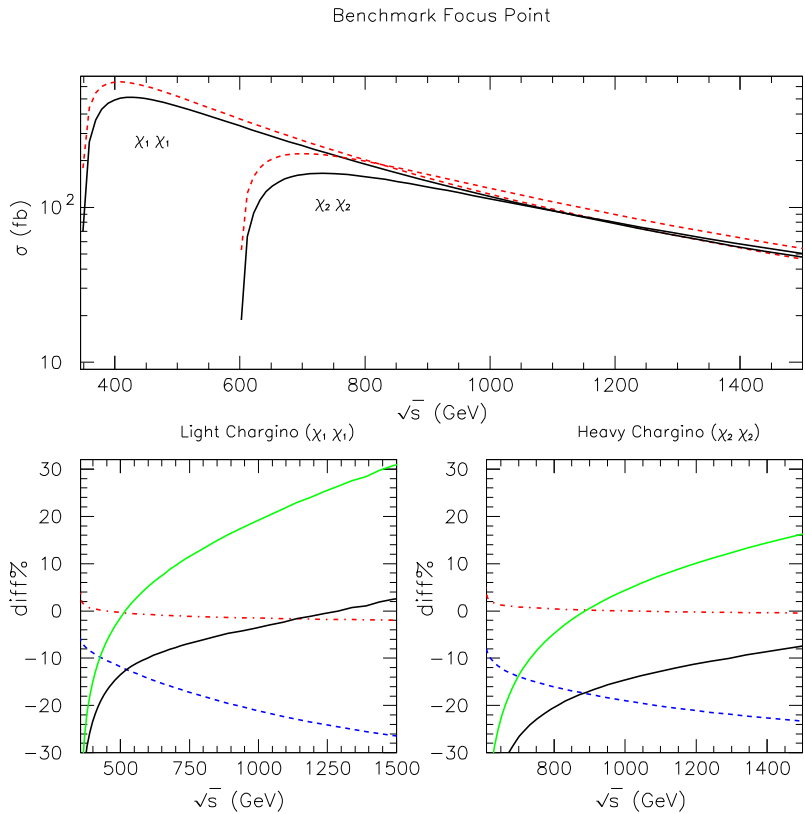}
\end{center}
\caption{Unpolarized   total   cross   section   as  a   function   of
  $\sqrt{s}$. The different curves correspond to full NLO corrected 
(black solid), MSSM (blue dashes), QED (red dot-dash), and  $\ell\ell$ 
(green solid).
\label{focus_point_runs}}
\end{figure}

Of course, the magnitude of quantum corrections depends on the supersymmetric
benchmark chosen. We already mentioned that our working scenario is benchmark 
SPS2, also known as Focus Point. In Fig.~\ref{focus_point_runs} we  show the 
one-loop unpolarized total cross  section   as  a   function  of  the   
center  of   mass  energy $\sqrt{s}$. In  the upper plot we have the total 
cross section in the Born approximation (red dash line), and the NLO corrected 
total cross section (black solid line), for light and heavy charginos. We see 
that in this scenario NLO corrections  to the  total cross  section for light
charginos, change  sign  at  $\sqrt{s}\approx  1250$  GeV, while for heavy 
charginos the correction is always negative in region shown in the graph.
%
%
In the lower frames of Fig.~\ref{focus_point_runs} we show details of 
the NLO corrections, as in the previous figure.
The magnitude of the corrections rise sharply as the energy approaches 
the threshold, although this percentage correction acts on a total cross 
section that approaches zero in this limit. 

We note that the corrections in the case of Focus point SPS2 considered in
this paper, are considerable different from those of benchmark SPS1a. The 
MSSM correction (excluding the pure QED corrections) are consistently lower 
(negative) in the case of SPS2. The major difference comes form the l.l. part 
of the QED corrections which, at high energies, are nearly three times as 
large as the case of SPS1a. This substantial difference is in turn dominated 
by a difference in the initial radiation correction to the sneutrino exchange 
part of the amplitude. For SPS2 the sneutrino mass is much larger than for 
SPS1a, so that its contribution to the  tree-level  differential cross-section 
is very insensitive to scattering angle, and negligible in front of s-channel
contributions.
On the other hand, sneutrino contribution in SPS1a scenario is large, and 
interferes destructively with s-channel contributions. All in all this leads 
to a light chargino total correction for the SPS2 point which is positive for 
$\sqrt{s} \, > \, $ 1250 GeV, whereas for SPS1a, the total correction remains 
negative.

%
\subsection{Polarized Differential Cross Sections}
%
In  this   section  we  show   our  results  for   one-loop  corrected
differential  cross  sections for  the  production  of light charginos  with
definite  helicity  from polarized  electron  positron collisions, in 
the Focus Point SPS2 scenario.  We
choose to  show our results for  the differential cross  sections as a
function  of  $\chi^+_1$ transverse  momentum  $p_T$ (rather than  the
scattering angle) defined as,
\begin{equation}
p_T^2=\frac{4}{s}(p_1\cdot k_2)(p_2\cdot k_2)-m_{\chin_1^+}^2
\end{equation}
where the  center of mass energy  is $\sqrt{s}=1000$ GeV.  In the case
there   is  no   photon  in   the   final  state,   this  reduces   to
$p_T=\sqrt{s}\,v\sin\theta/2$, where  $v$ and $\theta$  are $\chi^+_1$
velocity and scattering angle.

The differential cross  section we are working with  is related to the
differential  cross  section as  a  function  of  $\cos\theta$ in  the
following way,
\begin{equation}
\frac{d\sigma}{dp_T}=\frac{2\tan\theta}{\sqrt{s}\,v}
\frac{d\sigma}{d\cos\theta}
\end{equation}
which   partly   explains  the   tendency   for   the  cross   section
${d\sigma}/{dp_T}$ to  grow with  $p_T$, as we  go from  $\theta=0$ to
$\theta=\pi/2$.

In the following six figures  we show quantum corrections to polarized
chargino  production  cross  section   as  a  function  of  transverse
momentum. All of  these figures have the same  structure and differ in
the chargino helicities and the electron polarization.
In all  plots we show  $p_T>100$ GeV because  it is the  most interesting
region,   with   larger  cross   sections   and   better  chances   of
differentiation from background.

%
\begin{figure}[!h]
\begin{center}
\includegraphics[angle=0,width=9cm]{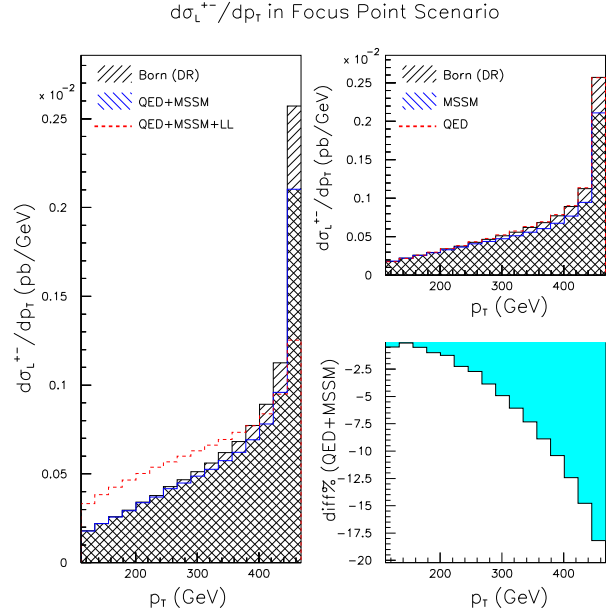}
\end{center}
\caption{$d\sigma_L^{+-}/dp_T$
  as  a  function  of  $p_T$  in the Focus  Point  scenario  within
  mSUGRA.\label{focus_point_Lpm}}
\end{figure}
%
In Fig.~\ref{focus_point_Lpm} we have $d\sigma_L^{+-}/dp_T$, i.e., for
a  chargino  with positive  helicity,  an  antichargino with  negative
helicity, and a  left handed electron (right handed  positron). On the
left we  have the differential cross  section as a  function of $p_T$,
where we compare the  born approximation, the one-loop corrected cross
section including  MSSM and QED  corrections, and the  corrected cross
section  including   MSSM,  QED,  and   leading  logarithms.  QED+MSSM
corrections are moderate and negative, varying between -1\% and -18\%,
as can be  seen also in the lower right  frame. Leading logarithms are
very  large, and increase the differential  cross section even more,
except  at large $p_T$,  where $\ell\ell$  are negative.   The largest
positive  corrections  are   obtained  for  low  transverse  momentum,
$p_T\sim 100$  GeV, while the  largest negative corrections  appear at
$p_T\sim  440$  GeV.  This  choice  for the  chargino  helicities  and
electron polarization  gives the  highest values for  the differential
cross section, varying between 0.2  and 2.1 fb/GeV. In the upper right
frame we show the MSSM  and QED corrections separately from each other,
and  compared  with the  Born  approximation.   We  see that  the  QED
corrections (not including leading logarithms) are smaller than 1\%.

%
\begin{figure}[!h]
\begin{center}
\includegraphics[angle=0,width=9.3cm]{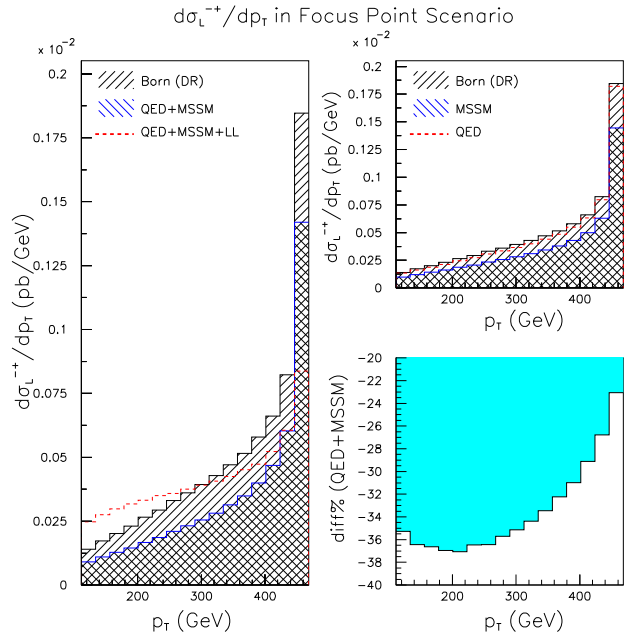}
\end{center}
\caption{$d\sigma_L^{-+}/dp_T$
  as  a  function  of  $p_T$  in the Focus  Point  scenario  within
  mSUGRA.\label{focus_point_Lmp}}
\end{figure}
%
In  Fig.~\ref{focus_point_Lmp}  we  have $d\sigma_L^{-+}/dp_T$,  which
corresponds to  the case with  both chargino helicities  reversed with
respect  to  the previous  case.  The  differential  cross section  is
comparable, varying between 0.1 and 1.4  fb/GeV, as we see in the left
frame. In the  lower right frame we observe  that QED+MSSM corrections
are larger  in magnitude  than the previous  case, and  also negative.
Note  that here the  relative magnitude of the 
corrections  decrease with
$p_T$, as opposed to the  previous case, where they increase.  Another
difference with the previous case  is that QED corrections are larger,
but  smaller  in magnitude  than  MSSM  corrections,  as seen  in  the
upper-right frame. This is a case where QED quantum corrections should
not  be neglected. The  Leading Logarithms  are usually  positive, but
become negative at large $p_T$, owing to the fact that at large $p_T$,
$x$ is  required to  be larger than  $4p_T^2/s$, thereby  reducing the
$x$-dependent phase-space factor $\Phi_2(x)$ in eq.(\ref{lleq}).
All the above differences between the corrections to $\sigma_L^{+-}$
and $\sigma_L^{-+}$ are interesting manifestations of the parity violation 
properties of the MSSM.

%
\begin{figure}[!h]
\begin{center}
\includegraphics[angle=0,width=9.3cm]{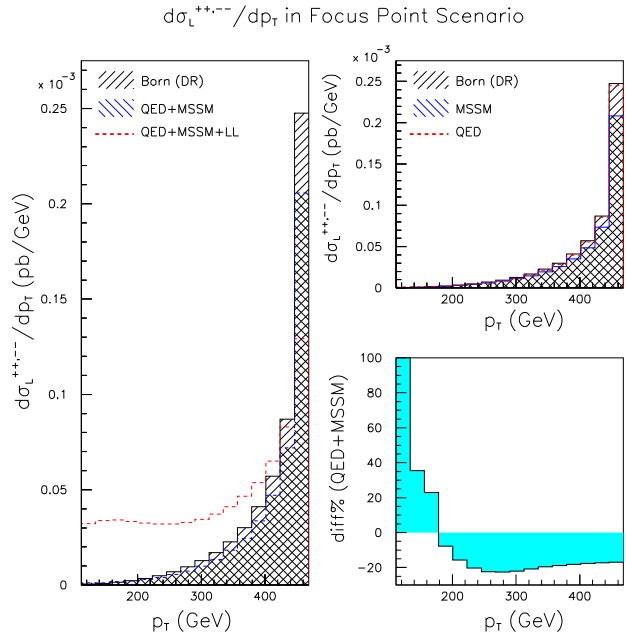}
\end{center}
\caption{$d\sigma_L^{++,--}/dp_T$
  as  a  function  of  $p_T$  in the  Focus  Point  scenario  within
  mSUGRA.\label{focus_point_Lpp}}
\end{figure}
%
In  Fig.~\ref{focus_point_Lpp}  we  see  the equal  chargino  helicity
production cross section. CP invariance, which we assume, enforces the
equality  of $d\sigma_L^{++}/dp_T$  and $d\sigma_L^{--}/dp_T$,  and it
serves  as a check  of our  calculations. This  cross section  is much
smaller, varying from less than 0.001 fb/GeV at low $p_T$ up to almost
0.2  fb/GeV at  large  $p_T$. In  the  lower-right frame  we see  that
QED+MSSM quantum corrections are negative at large $p_T$, and positive
at low  $p_T$, although  in that case  they are correcting  
a cross section which is already very small
at tree-level. From the upper-right frame we learn that QED
corrections are smaller than  MSSM corrections, however they cannot be
neglected  at   medium  values  for  $p_T$.   Leading  logarithms  are
comparatively large,  but in absolute  terms they are smaller  than in
the different helicity cases.

%
\begin{figure}[!h]
\begin{center}
\includegraphics[angle=0,width=9.3cm]{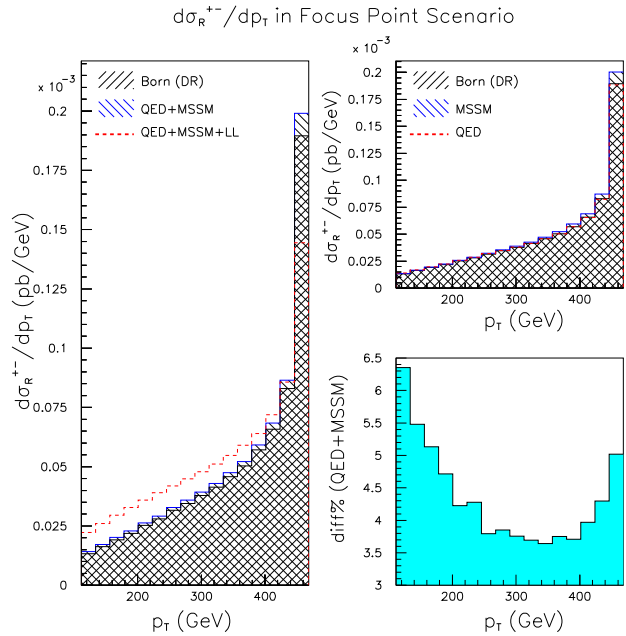}
\end{center}
\caption{$d\sigma_R^{+-}/dp_T$
  as  a  function  of  $p_T$  in the Focus  Point  scenario  within
  mSUGRA.\label{focus_point_Rpm}}
\end{figure}
%
In the  next three figures  we plot differential cross  sections where
right handed electrons collide with left handed positrons. These cross
sections  are  noticeably  smaller  than  the case  with  left  handed
electrons. The  reason is an  accidental cancellation between  $Z$ and
photon contributions  in the amplitude already  present at tree-level,
as  it was  explained  before. In  Fig.~\ref{focus_point_Rpm} we  have
$d\sigma_R^{+-}/dp_T$, which  varies from 0.01 to 0.2  fb/GeV as $p_T$
grows from  100 GeV up  to 468.8 GeV,  which is the maximum  value for
$p_T$.  QED+MSSM corrections vary  slowly between $3\%$-$6\%$, as seen
in the lower-right frame.  Nevertheless, from the upper-right frame we
see that most of these corrections come from MSSM loops and QED can be
neglected. In  addition, leading  logarithms can be  seen in  the left
frame,  and they are  large and  positive with  the exception  of very
large $p_T$ where they can be negative.

%
\begin{figure}[!h]
\begin{center}
\includegraphics[angle=0,width=9.3cm]{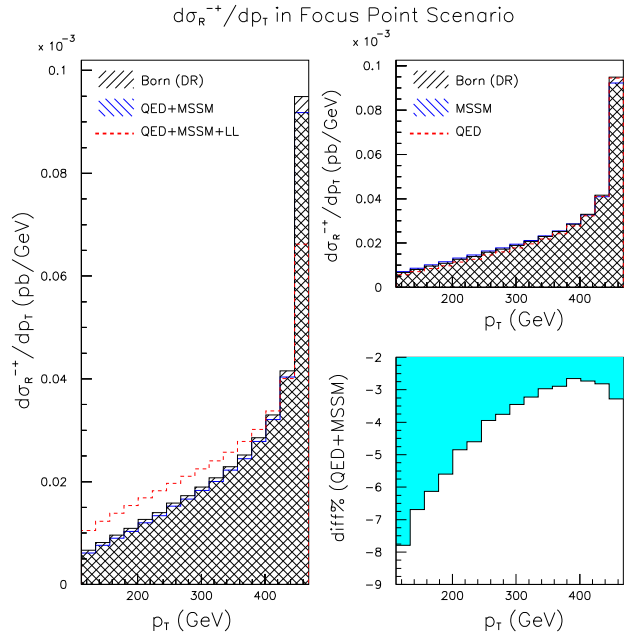}
\end{center}
\caption{$d\sigma_R^{-+}/dp_T$
  as  a  function  of  $p_T$  in the  Focus  Point  scenario  within
  mSUGRA.\label{focus_point_Rmp}}
\end{figure}
%
In Fig.~\ref{focus_point_Rmp} we  have $d\sigma_R^{-+}/dp_T$, which is
a  factor  1/2 smaller  than  the  previous  one. Quantum  corrections
corresponding to  QED+MSSM are larger, between $-3\%$  and $-8\%$, and
interestingly  enough they  have  different sign  than  the case  with
opposite chargino  helicities, thus  they cancel and  not add  to each
other as in  the case for left handed  electrons.  Nevertheless, these
large corrections for right handed electrons act on a 10 times smaller
cross section  compared with left  handed electrons. If  electrons are
not $100\%$ polarized (as expected) the pure right handed cross
section will be diluted by the  left handed cross section. We see from
the upper-right frame  that the QED corrections are  of the same order
than  MSSM ones, and  non-negligible.   Leading logarithms  are large,
positive for small $p_T$ and negative for high $p_T$.

%
\begin{figure}[!h]
\begin{center}
\includegraphics[angle=0,width=9.3cm]{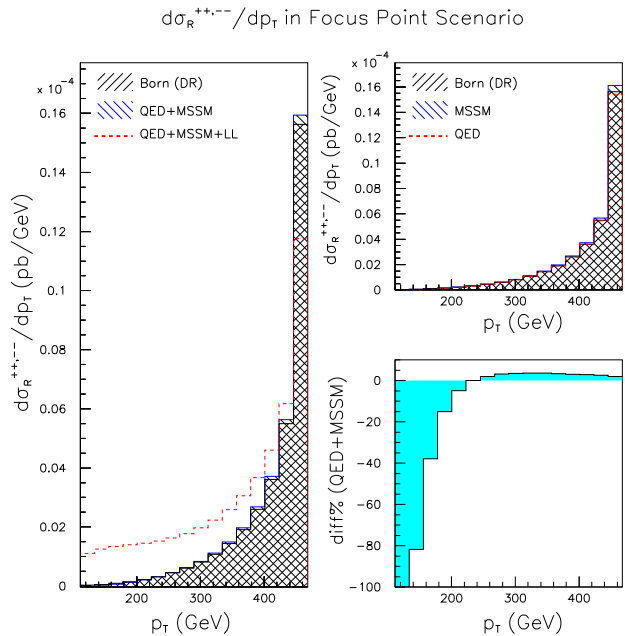}
\end{center}
\caption{$d\sigma_R^{++,--}/dp_T$
  as  a  function  of  $p_T$  in the Focus  Point  scenario  within
  mSUGRA.\label{focus_point_Rpp}}
\end{figure}
%
Finally,  in  Fig.~\ref{focus_point_Rpp} we  have  the equal  chargino
helicities    production   cross   section    $d\sigma_R^{++}/dp_T   =
d\sigma_R^{--}/dp_T$ for  the case of right handed  electrons and left
handed positrons. This cross  section is extremely small, never bigger
than 0.02 fb/GeV.  Quantum corrections are very large  at small $p_T$,
but with  cross sections so  small that the  chance to be  observed is
minimal. We include it for completness.
%
\section{Conclusions}
%
We have  calculated the complete  one-loop quantum corrections  to the
differential  cross  section   for  chargino  production  in  electron
positron collisions, at 1 TeV  center of mass energy, relevant for the
future International Linear Collider. In the analysis we have included
the  electron  and positron  polarization,  as  well  as the  chargino
helicities.

We   organize  the  non-QED   corrections  calculating   the  helicity
amplitudes,   and  regularizing   the  ultraviolet   divergences  with
Dimensional  Regularization. Our  renormalization  scheme includes  on
shell masses  and $\overline{DR}$  running couplings. In  addition, we
use the dipole  subtraction formalism in Bremsstrahlung contributions
to cancel  analytically soft and collinear  divergences, with infrared
divergences  from virtual  QED  corrections.  The  well known  leading
logarithms remain, and  we isolate them in order  to better understand
quantum corrections.

We show numerical results for the Focus Point scenario in mSUGRA.  The
dipole subtraction  formalism is designed to  handle analytically the
divergences,   therefore  the   use  of   infrared   cut-offs  becomes
unnecessary,  making  the  calculation  numerically  stable.  We  find
important  quantum corrections  for differential  cross  sections with
definite chargino  helicities. 
The substantial differences in the NLO corrections depending on the 
helicities of the charginos and of the initial leptons, show that the 
rich structure of the MSSM - in particular the parity violating 
properties - can be investigated, firstly by polarization of the incoming electron/proton beam and secondly by determining (through the angular 
distributions of the decay products) the helicities of the charginos 
produced \cite{gudrid,Bernreuther:2001rq}.
Since charginos are  unstable, the full
potential  of  this work  will  be  fulfilled  when similar  radiative
corrections  are  included  for  the  chargino  decay,  and  the  spin
correlations  between  production   and  decay  are  evaluated.  Large
corrections  of the  order of  -18\% and  -35\% are  obtained  for the
largest      cross       sections,      $d\sigma_L^{-+}/dp_T$      and
$d\sigma_L^{+-}/dp_T$. They are large enough to making it necessary to
include them in any analysis for the ILC.
%
\acknowledgments
\label{subsub:ackn}
%
M.A.D. was supported partly by Anillo "Centro de Estudios Subatomicos"
grant. M.A.R. was supported by Fondecyt grant No. 3090069. D.A.R.  was
supported    partly   Comisi\/{o}n   Nacional    de   Investigci\'{o}n
Cient\'{i}fica  Y Tecnol\'{o}gica  (CONICYT) and  wishes to  thank the
Departamento de  Fisica, Pontificia Universidad  Cat\'{o}lica de Chile
for its warm hospitality.
%
%
%


\begin{thebibliography}{99}

\bibitem{Haber:1984rc}
H.~E.~Haber and G.~L.~Kane,
 Phys.\ Rept.\  {\bf 117}, 75 (1985).


\bibitem{Abbiendi:2003sc}
  G.~Abbiendi {\it et al.}  [OPAL Collaboration],
  Eur.\ Phys.\ J.\  C {\bf 35}, 1 (2004)
  [arXiv:hep-ex/0401026].

\bibitem{Abdallah:2003xe}
  J.~Abdallah {\it et al.}  [DELPHI Collaboration],
  Eur.\ Phys.\ J.\  C {\bf 31}, 421 (2004)
  [arXiv:hep-ex/0311019].

\bibitem{Heister:2002mn}
  A.~Heister {\it et al.}  [ALEPH Collaboration],
  Phys.\ Lett.\  B {\bf 533}, 223 (2002)
  [arXiv:hep-ex/0203020].

\bibitem{Acciarri:1999km}
  M.~Acciarri {\it et al.}  [L3 Collaboration],
  Phys.\ Lett.\  B {\bf 472}, 420 (2000)
  [arXiv:hep-ex/9910007].

\bibitem{Aaltonen:2008pv}
  T.~Aaltonen {\it et al.}  [CDF Collaboration],
  Phys.\ Rev.\ Lett.\  {\bf 101}, 251801 (2008)
  [arXiv:0808.2446 [hep-ex]].

\bibitem{Abazov:2009zi}
  V.~M.~Abazov {\it et al.}  [D0 Collaboration],
  arXiv:0901.0646 [hep-ex].


\bibitem{Polesello:2004aq}
  G.~Polesello,
  J.\ Phys.\ G {\bf 30} (2004) 1185.

\bibitem{Blumenschein:2005ms}
  U.~Blumenschein,


\bibitem{Kalinowski:1998yn}
  J.~Kalinowski,
  arXiv:hep-ph/9905558.


\bibitem{Pierce:1993gj}
D.~Pierce and A.~Papadopoulos,
Phys.\ Rev.\  D {\bf 50}, 565 (1994)

\bibitem{Eberl:2001eu}
  H.~Eberl, M.~Kincel, W.~Majerotto and Y.~Yamada,
  Phys.\ Rev.\  D {\bf 64}, 115013 (2001)
  [arXiv:hep-ph/0104109].

\bibitem{Fritzsche:2002bi}
  T.~Fritzsche and W.~Hollik,
  Eur.\ Phys.\ J.\  C {\bf 24}, 619 (2002)
  [arXiv:hep-ph/0203159].

\bibitem{Schofbeck:2007ib}
  R.~Schofbeck and H.~Eberl,
  Eur.\ Phys.\ J.\  C {\bf 53}, 621 (2008)
  [arXiv:0706.0781 [hep-ph]].





\bibitem{Diaz:1997kv}
  M.~A.~Diaz, S.~F.~King and D.~A.~Ross,
  Nucl.\ Phys.\  B {\bf 529}, 23 (1998)
  [arXiv:hep-ph/9711307].

\bibitem{Ellis:1998jk}
J.~R.~Ellis, T.~Falk, G.~Ganis, K.~A.~Olive and M.~Schmitt,
Phys.\ Rev.\  D {\bf 58}, 095002 (1998)

\bibitem{Oller:2005xg}
W.~Oller, H.~Eberl and W.~Majerotto,
Phys.\ Rev.\  D {\bf 71}, 115002 (2005)

\bibitem{Kilian:2006cj}
  W.~Kilian, J.~Reuter and T.~Robens,
  Eur.\ Phys.\ J.\  C {\bf 48}, 389 (2006)

\bibitem{Robens:2006np}
  T.~Robens,
  arXiv:hep-ph/0610401.

\bibitem{Blank:2000uc}
  T.~Blank and W.~Hollik,
  arXiv:hep-ph/0011092.

\bibitem{Kiyoura:1998yt}
  S.~Kiyoura, M.~M.~Nojiri, D.~M.~Pierce and Y.~Yamada,
  Phys.\ Rev.\  D {\bf 58}, 075002 (1998)
  [arXiv:hep-ph/9803210].


\bibitem{Hao:2006df}
  S.~Hao, H.~Liang, M.~Wen-Gan, Z.~Ren-You, J.~Yi and G.~Lei,
  Phys.\ Rev.\  D {\bf 73}, 055002 (2006)
  [arXiv:hep-ph/0602089].


\bibitem{Fujimoto:2007bn}
  J.~Fujimoto, T.~Ishikawa, Y.~Kurihara, M.~Jimbo, T.~Kon and M.~Kuroda,
  Phys.\ Rev.\  D {\bf 75}, 113002 (2007).

\bibitem{Zhang:2001rd}
  R.~Y.~Zhang, W.~G.~Ma and L.~H.~Wan,
  J.\ Phys.\ G {\bf 28}, 169 (2002)
  [arXiv:hep-ph/0111124].


\bibitem{Diaz:2001vm}
  M.~A.~Diaz and D.~A.~Ross,
  JHEP {\bf 0106}, 001 (2001)
  [arXiv:hep-ph/0103309].

\bibitem{Diaz:2000hi}
  M.~A.~Diaz, S.~F.~King and D.~A.~Ross,
  Phys.\ Rev.\  D {\bf 64}, 017701 (2001)
  [arXiv:hep-ph/0008117].

\bibitem{Diaz:2002rr}
  M.~A.~Diaz and D.~A.~Ross,
  arXiv:hep-ph/0205257.

\bibitem{Baer:2002bb}
  H.~Baer, M.~A.~Diaz, M.~A.~Rivera and D.~A.~Ross,
  arXiv:hep-ph/0210444.


\bibitem{gudrid}
G. Moortgat-Pick, H. Fraas, {\sl Phys. Rev. D} {\bf 59}, 015016
(1999); G. Moortgat-Pick, H. Fraas, A. Bartl, W. Majerotto, {\sl
Eur. Phys. J.} {C7}, 113 (1999); G. Moortgat-Pick, H. Fraas, A.
Bartl, W. Majerotto, {\sl Eur. Phys. J.} {\bf C9}, 521 (1999),
Erratum-ibid. {\bf C9}, 549 (1999); G. Moortgat-Pick, A. Bartl, H.
Fraas, W. Majerotto, {\sl Eur. Phys. J.} {\bf C18}, 379 (2000).


\bibitem{Dittmaier:1999mb}
S.~Dittmaier,
Nucl.\ Phys.\  B {\bf 565}, 69 (2000)


\bibitem{Yao:2006px}
  W.~M.~Yao {\it et al.}  [Particle Data Group],
  J.\ Phys.\ G {\bf 33} (2006) 1.


\bibitem{AguilarSaavedra:2005pw}
  J.~A.~Aguilar-Saavedra {\it et al.},
  Eur.\ Phys.\ J.\  C {\bf 46}, 43 (2006)
  [arXiv:hep-ph/0511344].


\bibitem{Abbiendi:2000hu}
  G.~Abbiendi {\it et al.}  [OPAL Collaboration],
  Eur.\ Phys.\ J.\  C {\bf 19}, 587 (2001)
  [arXiv:hep-ex/0012018];
  P.~Abreu {\it et al.}  [DELPHI Collaboration],
  Eur.\ Phys.\ J.\  C {\bf 16} (2000) 371;
  M.~Acciarri {\it et al.}  [L3 Collaboration],
  Eur.\ Phys.\ J.\  C {\bf 16}, 1 (2000)
  [arXiv:hep-ex/0002046];
  R.~Barate {\it et al.}  [ALEPH Collaboration],
  Eur.\ Phys.\ J.\  C {\bf 14}, 1 (2000).

\bibitem{Bloch:1937pw}
F.~Bloch and A. Nordsieck,
Phys.\ Rev.\  {\bf 52}, 54 (1937).

\bibitem{Kinoshita:1962ur}
T.~Kinoshita,
J.\ Math.\ Phys.\  {\bf 3}, 650 (1962).

\bibitem{Lee:1964is}
T. D. Lee and M. Nauenberg,
Phys.\ Rev.\  {\bf 133}, B1549 (1964).

\bibitem{Baer:1999sp}
  H.~Baer, F.~E.~Paige, S.~D.~Protopopescu and X.~Tata,
  arXiv:hep-ph/0001086.

\bibitem{Vermaseren:2000nd}
J.~A.~M.~Vermaseren,
\textsf{``New Features Of FORM,''}
arXiv:math-ph/0010025.


\bibitem{Allanach:2002nj}
  B.~C.~Allanach {\it et al.},
in {\it Proc. of the APS/DPF/DPB Summer Study on the Future of Particle Physics (Snowmass 2001) } ed. N.~Graf,
  Eur.\ Phys.\ J.\  C {\bf 25}, 113 (2002).

\bibitem{Bernreuther:2001rq}
  W.~Bernreuther, A.~Brandenburg, Z.~G.~Si and P.~Uwer,
  Phys.\ Rev.\ Lett.\  {\bf 87}, 242002 (2001)
  [arXiv:hep-ph/0107086].

\end{thebibliography}
\end{document}